\documentclass{ws-procs961x669}            
\usepackage[normalem]{ulem}
\usepackage{hyperref}
\usepackage{xcolor}
\usepackage{soul}

\begin{document}
\title{Probing dense matter physics with transiently-accreting neutron stars: the case of source MXB~1659-29}

\author{Melissa Mendes$^{*}$, Andrew Cumming, Charles Gale}

\address{Department of Physics, McGill University,\\
3600 rue University, Montreal, QC, H3A 2T8, Canada\\
$^*$E-mail: melissa.mendessilva@mail.mcgill.ca}

\author{Farrukh J. Fattoyev}
\address{Department of Physics, Manhattan College, Riverdale, NY 10471, USA}

\begin{abstract}

Recent observational data on transiently-accreting neutron stars has unequivocally shown fast-cooling sources, such as in the case of neutron star MXB~1659-29. Previous calculations have estimated its total neutrino luminosity and heat capacity, as well as suggested that direct Urca reactions take place in $1 \%$ of the volume of the core. In this paper, we reproduce the inferred luminosity of this source with detailed models of equations of state (EOS) and nuclear pairing gaps. We show that three superfluidity gap models are inconsistent with data for all EOS and another three are disfavoured because of fine tuning arguments. We also calculate the total heat capacity for all constructed stars and show that independent observations of mass and luminosity could set constraints on the core superfluidity of a source as well as the density slope of the symmetry energy, $(L)$. This is an important step towards defining a universal equation of state for neutron stars and therefore, towards a better understanding of the phase diagram of asymmetric matter at high densities.
\end{abstract}

\keywords{symmetry energy; neutron star composition; direct Urca; transiently-accreting neutron stars}

\bodymatter

\section{Introduction}

The composition of the core of neutron stars is still unknown. Their extreme densities make it impossible for similar conditions to be reproduced experimentally in laboratories, and their strongly-coupled hadronic nature represents a serious challenge for first-principles calculations. Therefore, there is a lot of uncertainty when it comes to the particle content and matter organization in neutron star cores. Even in the simplest model of neutron stars made only of neutrons, protons, electrons and muons, there is no agreement on the star's proton fraction -- which determines electron and muon fractions, by the charge neutrality requirement -- or how to describe proton superconductivity and neutron superfluidity, both expected to happen in matter at high densities \cite{Page:2013hxa}. The proton fraction is important because it determines whether neutrino cooling occurs via modified Urca or direct Urca reactions, which differ by orders of magnitude in emissivity.

Although BCS theory with polarization corrections\,\cite{Bardeen1957,Gezerlis:2007fs} can describe superfluidity and superconductivity in a low density medium, at high densities, in-medium effects prevent an accurate description of neutron and proton pairing. This uncertainty forbids a precise determination of the critical temperatures as a function of density, that is, the gap amplitude and width of the pairing gap in neutron star cores. Nonetheless, a wide range of theoretical calculations of superfluidity at high densities exist, which have been summarized in the functional analytic form in Ref. \citenum{Hoetal}. Its parameters mimic the effect of in-medium interactions, and thus define a range of possible models for nucleon pairing at high densities, each one with their specific amplitudes and widths.

These different gap models will predict different stellar cooling rates, as shown in Ref. \citenum{Hoetal}, because both superfluidity and superconductivity suppress neutrino production through direct or modified Urca reactions \cite{Yakovlev}. Furthermore, Cooper pair formation produces pair-breaking-formation (PBF) cooling reactions, which enhance the star's cooling when its temperature is close to the superfluid critical temperature \cite{Yakovlev}. The suppression or enhancement of neutrino emission due to superfluidity depends on density, and therefore location within the neutron star core. By comparing these predictions with the star's observed luminosity, one can exclude or favor specific gap models.

Transiently-accreting neutron stars are ideal systems to perform these observations. These stars periodically accrete matter from a companion star, in cycles of accretion outbursts followed by periods of quiescence, in which accretion halts \cite{review_accreting, LattimerComment}. By observing X-ray emissions after the accretion period, the surface temperature of the star can be found. Combining that information with the mass accretion rate, one can obtain a consistent estimate of the star's neutrino luminosity and determine whether fast cooling reactions (such as direct Urca) or slow cooling reactions (such as modified Urca) are taking place in the neutron star's core \cite{LattimerComment}.

The source MXB~1659-29 is particularly interesting because it has shown more than one accretion-quiescence cycle, which allows for repeated measurements of accretion rates during outburst, luminosity and temperature during quiescence. Analysis of the energy balance between accretion-driven crustal heating in outburst and neutrino cooling during quiescence leads to the conclusion that this source is undergoing fast cooling processes in its core \cite{Brownetal, LattimerComment}. Furthermore, it has been suggested in Ref. \citenum{Brownetal} that measuring the decrease of external temperature of the  neutron star MXB~1659-29 over an interval of ten years, combined with the observation of its neutrino luminosity, could set constraints on its total heat capacity, and therefore on its composition. This analysis would represent another avenue for finding, for example, the relative contribution of leptons to the total heat capacity of a specific source based on direct observations, which could set further constraints on its superfluidity.

Other transiently accreting sources such as KS~1731-260 have inferred neutrino cooling luminosities inconsistent with fast cooling, implying that they have slow cooling processes operating in their cores \cite{Brownetal2017}. To explain this data, one should consider a combination of equation of state (EOS) describing a star's particle content, and a collection of nuclear pairing gap models that accommodate both fast and slow cooling processes. In this paper, we create detailed realistic scenarios for MXB~1659-29, to determine how the observed neutrino cooling rate of that source constrains the input physics. We use gap models described in Ref. \citenum{Hoetal}, and a family of relativistic mean-field (RMF) EOS \cite{Reed:2021nqk} based on FSUGold2, first described in Ref. \citenum{FSUGoldReference}. Their particularities are discussed in section \ref{formalism}. In section \ref{results}, we investigate the agreement of the calculated luminosities with data and we verify the suggestion made in Ref. \citenum{Brownetal} that direct Urca processes occur in approximately $1 \%$ of the volume of the star's core. Finally, in section \ref{discussion}, we discuss which gap models accurately describe this scenario and compute predictions for the neutron star heat capacity in each case.

\section{Formalism}\label{formalism}

To determine the particle composition of the core of neutron stars, we use a family of equations of state (EOS) that are developed from the RMF model FSUGold2 \cite{FSUGoldReference}. The original FSUGold2 model was created to reproduce the ground-state properties of finite nuclei such as binding energy and charge radii, their monopole response, and the maximum observed neutron star mass. In particular, the original model predicts both a stiff symmetry energy and a soft equation of state for symmetric nuclear matter, that is, matter with equal number of protons and neutrons. The symmetry energy is an essential ingredient of the EOS that strongly impacts the structure, dynamics, and composition of neutron stars and has received considerable attention over the last decade\,\cite{Horowitz:2014bja, Li:2014Sym}. Customarily, one expands the total energy per nucleon $E(\rho, \alpha)$ at zero temperature---where $\rho = \rho_{\rm n} + \rho_{\rm p}$ is the total baryon density and $\alpha=(\rho_{\rm n} - \rho_{\rm p})/\rho$ is the neutron-proton asymmetry parameter---around the energy of isospin symmetric matter with $\alpha = 0$,
\begin{align}
E(\rho, \alpha) = E_{\rm SNM}(\rho) + E_{\text {\rm sym}}(\rho) \cdot \alpha^{2}+\mathcal{O}\left(\alpha^{4}\right) 
\label{EnNucleon} \,
\end{align}
where  $E_{\rm SNM}(\rho) \equiv E(\rho, 0)$ is the energy per nucleon in symmetric nuclear matter (SNM) and $E_{\text {\rm sym}}(\rho)$ the symmetry energy, which represents a correction of second order in parameter $\alpha$, to the symmetric limit. No odd powers of $\alpha$ appear in the expansion above because the nuclear force is assumed to be isospin symmetric. To characterize the behavior of both $E_{\rm SNM}(\rho)$ and $E_{\text {\rm sym}}(\rho)$ near the nuclear saturation density $\rho_{\rm sat} \approx 0.15$ fm$^{-3}$, one can further expand  these quantities in a Taylor series\cite{Piekarewicz:2008nh},
\begin{align}
& E_{\rm SNM}(\rho) = B + \frac{1}{2} K x^2 + \cdots 
\label{SNM}\,\\
& E_{\rm sym}(\rho) = J + L x+ \frac{1}{2} K_{\rm sym} x^2 + \cdots \label{SymmetryEnergy}
\end{align}
where $x = (\rho - \rho_{\rm sat})/3\rho_{\rm sat}$ is a dimensionless parameter that quantifies the deviations of the density from its value at saturation, $B$ is the energy per nucleon and $K$ is the incompressibility coefficient of SNM. Similarly, $J$ and $K_{\rm sym}$ represent the symmetry energy and the incompressibility coefficient of symmetry energy at saturation density and serve as corrections to the binding energy and incompressibility. Unlike in SNM, whose pressure vanishes at saturation density, the slope of the symmetry energy $L$, and consequently, pressure of pure neutron matter, do not vanish at $\rho_{\rm sat}$. 

Since $L$ is poorly constrained experimentally, we generated a family of the FSUGold2 models that are identical in their predictions for the EOS of SNM but vary in their predictions for the symmetry energy. It is well known that, at a sub-saturation density of $\rho \approx 0.1\ \mathrm{fm}^{-3}$, which represents an average value between the central and surface densities, the symmetry energy is well constrained
by the binding energy of heavy nuclei with a significant
neutron excess\cite{Horowitz:2000xj, Furnstahl:2001un, Zhang:2013wna}. By fixing this value of the symmetry energy, we obtain a family of models with differing $(J,L)$ values that are identical in their predictions for the ground state properties of finite nuclei but predict a range of the neutron skin thicknesses and neutron star radii consistent with the current experimental and observational data\cite{Adhikari:2021phr, Riley:2019yda, Miller:2019cac, Abbott:PRL2017, Abbott:2018exr}.

\begin{figure}[ht]
	\begin{center}
	\includegraphics[width=5.2 in]{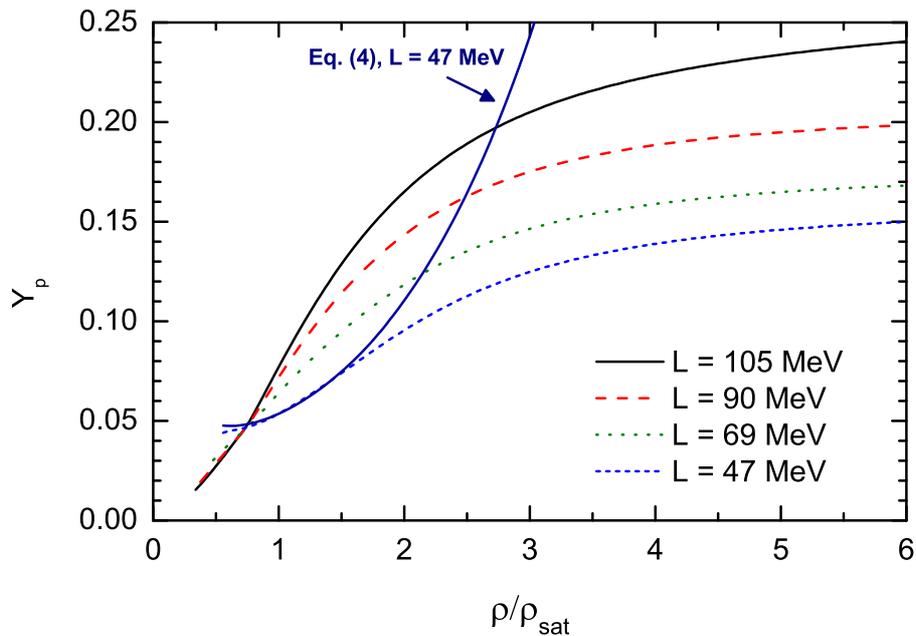}
	\end{center}
	\caption{Proton fraction $Y_{\rm p}$ versus baryon density $\rho$ normalized by saturation density for EOS in the FSUGold2 family. Different curves correspond to EOS with different $L$ values, in $\mathrm{MeV}$. The blue solid curve shows the analytic approximation (Eq. \ref{protonfraction}).}
	\label{fig1}
\end{figure}

Cold-catalyzed matter in neutron stars, ie.~matter in the ground state with lowest energy per nucleon, is in a state of chemical equilibrium and is assumed to be electrically neutral. Therefore one must impose a charge neutrality and chemical equilibrium condition to obtain an EOS of neutron-star matter. We consider a minimal model in which neutrons, protons, electrons and muons are present inside neutron stars. Increasing the parameter $L$ leads to a larger symmetry energy at supersaturation densities. This in turn makes it difficult for protons to convert into neutrons and thus increases the proton fraction $Y_{\rm p} = \rho_{\rm p}/\rho$ in the innermost region of the star, as shown in Fig. \ref{fig1}. For matter composed of neutrons, protons and electrons, Ref. \citenum{constraints} shows this proportionality explicitly by writing the beta equilibrium condition as a function of the proton fraction, such that for low proton fraction,
\begin{equation}\label{protonfraction}
Y_{\rm p} \simeq \frac{64}{3 \pi^{2} \rho_{\rm sat} \left(3x+1\right)}\left(\frac{J + L x + \frac{1}{2} K_{\rm sym} x^2}{\hbar c}\right)^{3}.
\end{equation}
As displayed in Fig. \ref{fig1}, this approximation is only valid up to about $1.5 \rho_{\rm sat}$, where proton fractions are small. Here we used the predicted bulk parameters of symmetry energy at saturation density $\rho_{\rm sat} = 0.15$ fm$^{-3}$, $J  =30.6$ MeV, $L  =47.0$ MeV and $K_{\rm sym} = 54.0$ MeV to generate the blue solid curve in Fig. \ref{fig1}.

The outer crust is described by the EOS from Ref. \citenum{crust1}, whereas the EOS for the inner crust is described in Ref. \citenum{crust2}. For the core we use the FSUGold2 family of EOS with different values of the slope of the symmetry energy $L$. We assume a non-rotating spherically-symmetric neutron star, and solve the Tolman-Oppenheimer-Volkoff (TOV) equations
\begin{align}
& \frac{\mathrm{d} P}{\mathrm{d} r} =-\frac{\mathcal{E}(r)}{c^2}\frac{G m(r)}{r^2} \left[1 + \frac{P(r)}{\mathcal{E}(r)}\right] \left[1+\frac{4 \pi r^{3} P(r)}{m(r) c^2}\right] \left[1-\frac{2Gm(r)}{c^2r}\right]^{-1} \ , \\[0.5em]
& \frac{\mathrm{d} m}{\mathrm{d} r} = 4 \pi r^2 \frac{\mathcal{E}(r)}{c^2} \ , \\[0.5em]
& \frac{\mathrm{d} \phi}{\mathrm{d} r} =- \frac{1}{\mathcal{E}(r)+P(r)} \frac{\mathrm{d} P}{\mathrm{d} r} \,
\end{align} 
where $m(r)$ is the mass within radius $r$, $P(r)$ is the pressure, $\mathcal{E}(r)$ is the energy density and $\phi(r)$ is the gravitational potential such that at the surface of the star, $r = R$ and $m=M$, the pressure vanishes, $P(R) = 0$ and $\phi(R)=\frac{1}{2} \ln (1-{2 G M}/{c^2R})$.

Since our goal is to reproduce the inferred neutrino luminosity of MXB~1659-29, we consider the fast cooling process of direct Urca only. If there are no muons participating, direct Urca cooling takes place through the reactions 
\begin{equation}
n \rightarrow p+e^{-}+\bar{\nu}_{e}, \quad p+e^{-} \rightarrow n+\nu_{e}.
\end{equation}
This process conserves momentum only if
\begin{equation}
k_{F n} \leq k_{F p}+k_{F e},
\end{equation}
which implies that for direct Urca reactions the proton fraction must exceed a threshold value 
\begin{equation}\label{condition}
Y_{p} \geq Y_{\mathrm{p\, dUrca}} = \frac{\left[\left(3 \pi^{2} \hbar^{3} \rho Y_{n}\right)^{1 / 3}-\left(3 \pi^{2} \hbar^{3} \rho_{e}\right)^{1 / 3}\right]^{3}}{3 \pi^{2} \hbar^{3} \rho},
\end{equation} 
as explained in Ref. \citenum{Yakovlev}.  Here, $k_{\mathrm{Fx}}$ are the Fermi momenta, a function of $\rho_x$, the number density for each species and the particle fraction $Y_{x}$. If there are no muons in the star at all, Eq. (\ref{condition}) can be simplified because of the star's charge neutrality condition $\rho_p = \rho_e$, hence $Y_{\mathrm{p\, dUrca}} =1/9 \approx 0.11$. However, for the EOS studied here, $Y_{\mathrm{p\, dUrca}} \neq 1/9$ because even when muons are not participating in direct Urca reactions, they are present, such that the charge neutrality condition becomes $\rho_p = \rho_e+ \rho_{\mu}$, thus, for our set of EOS, we have $0.131 \leq Y_{\mathrm{p\, dUrca}} \leq 0.138$.

\begin{figure}[ht]
\begin{center}
\includegraphics[width=0.98\columnwidth]{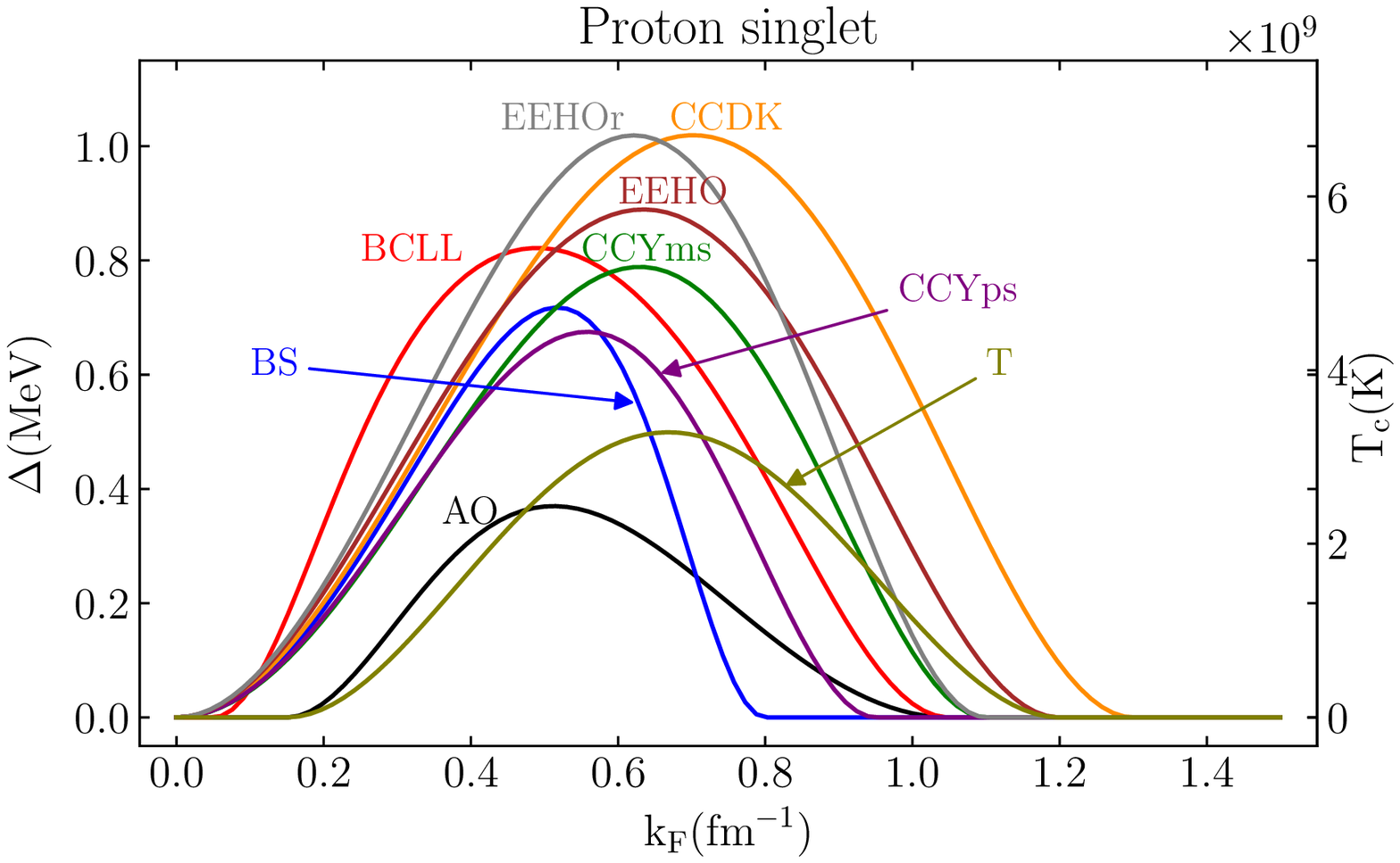}
\includegraphics[width=\columnwidth]{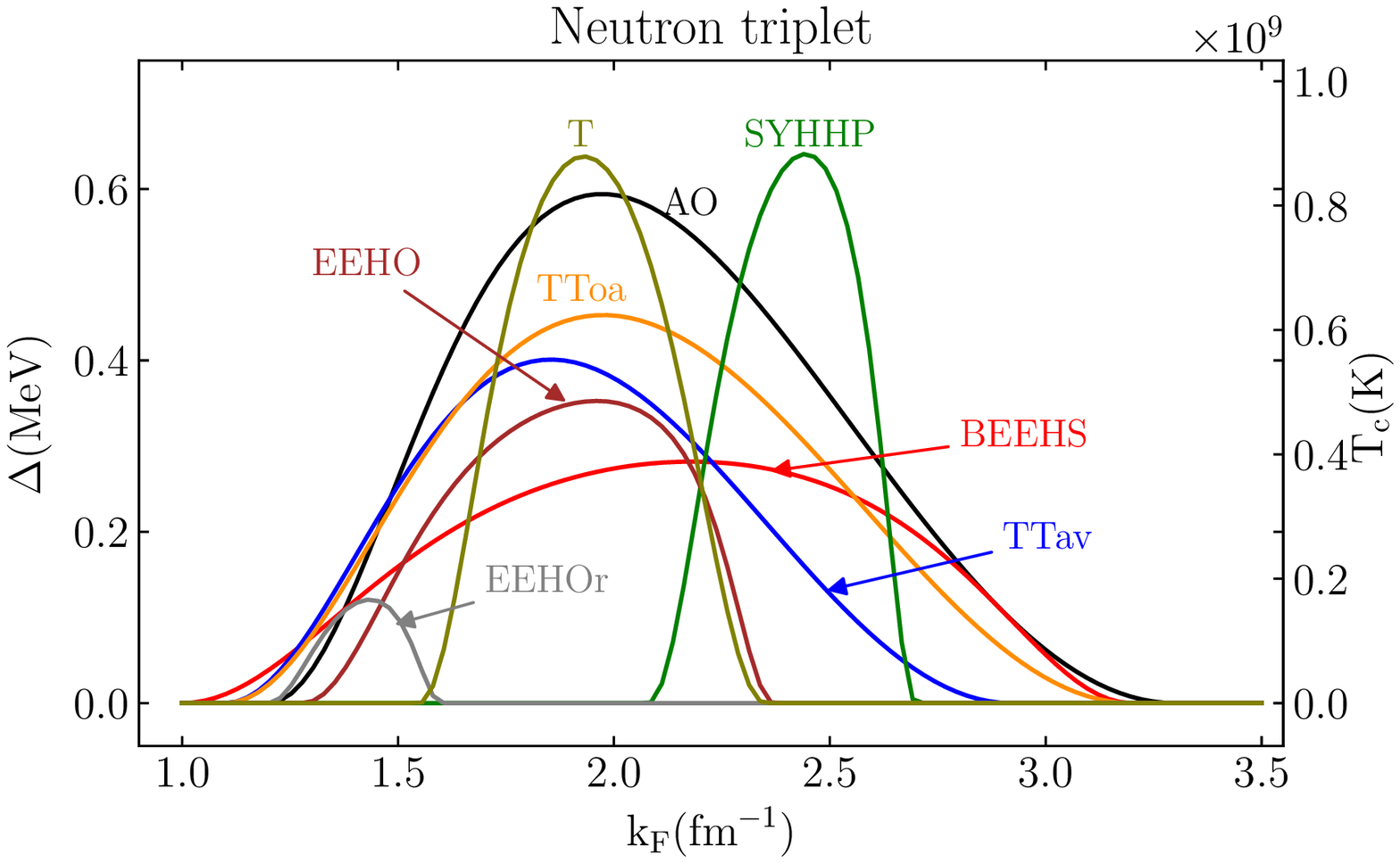}
\caption{Gap model parametrizations $\Delta \, (\mathrm{MeV})$ as a function of Fermi momenta $k_F \, (\mathrm{fm^{-1}})$ for all models studied here (see Ref. \citenum{Hoetal} and references therein).
	(a) Proton singlet superconductivity (b) Neutron triplet superfluidity.}
\label{parametrizations}
\end{center}
\end{figure}

The direct Urca neutrino luminosity is given by
\begin{equation}\label{integration}
L_{\nu_{\mathrm{dUrca}}} = \int_0^{R_\mathrm{core}}\frac{4 \pi r^2 \epsilon_{0}^{\mathrm{dUrca} ~ \mathrm{total}}}{\left(1-(2 G m(r)/c^ 2 r)\right)^{1/2}} dr,
\end{equation}
where the integral is over the neutron star core, and the local neutrino emissivity is\cite{Yakovlev} 
\begin{align}\label{emissivity}
\epsilon_{0}^{\mathrm{dUrca} ~ \mathrm{total}} &= \epsilon_{0}^{\mathrm{dUrca} ~ e^{-}}+ \epsilon_{0}^{\mathrm{dUrca} ~ \mu^{-}}\\
\epsilon_{0}^{\mathrm{dUrca} ~ e^{-}}&=\frac{457 \pi}{10080} G_{\mathrm{F}}^{2} \cos ^{2} \theta_{\mathrm{C}}\left(1+3 g_{\mathrm{A}}^{2}\right) \frac{m_{n}^{*} m_{\mathrm{p}}^{*} m_{e}^{*}}{h^{10} c^{3}}\left(k_{\mathrm{B}} T\right)^{6} \Theta_{\mathrm{npe}},
\end{align} 
where we used the weak coupling constant $G_{\mathrm{F}}= 1.436 10^{-62} \mathrm{J} \mathrm{m}^3 $, the Cabibbo angle $\theta_{\mathrm{C}} = 0.227$, the axial vector coupling constant $g_{\mathrm{A}}=-1.2601 \left(1-\frac{\rho}{4.15 (\rho_{0}+\rho)}\right)$. $\mathrm{M}_{x}^{*}$ represents the effective mass of species $x$ and $\Theta_{\mathrm{npe}}$ is a step function specifying the densities where direct Urca reactions can happen, respecting momentum conservation. Note that, in general, muons can participate in these reactions at high densities, so that
\begin{equation}\label{muonsincluded}
    n \rightarrow p+\mu^{-}+\bar{\nu}_{\mu^{-}}, \quad p+\mu^{-} \rightarrow n+\nu_{\mu^{-}}
\end{equation}
Their emissivities are
\begin{equation}
    \epsilon_{0}^{\mathrm{dUrca} ~ \mu^{-}}= \epsilon_{0}^{\mathrm{dUrca} ~ e^{-}}.
\end{equation}

Including superfluidity and superconductivity in the neutron star core model changes the neutrino luminosity calculations. The formation of Cooper pairs reduces the number of neutrons and/or protons available for participating in direct Urca reactions, therefore the rate $\epsilon^{\mathrm{dUrca}}$ is exponentially reduced, as described in  Ref. \citenum{Yakovlev}, according to
\begin{align}
    \epsilon^{\mathrm{dUrca}} &=\epsilon_{0}^{\mathrm{dUrca}} R\\
    R &=\exp \left(-v^{*}_{\mathrm{triplet}/\mathrm{singlet}}\right)\label{reduced1}\\
    \text{for }v_{\text {triplet }}^{*} &=2.376 \left(\mathrm{T}_{\mathrm{c}}/ \mathrm{T} \right), \text{where } \mathrm{T}_{\mathrm{c}}=0.1187 \Delta(k_{\mathrm{Fx}})\label{reduced2} \\
    \text{and }v_{\text {singlet }}^{*} &=1.764 \left(\mathrm{T}_{\mathrm{c}} / \mathrm{T} \right), \text{where } \mathrm{T}_{\mathrm{c}}=0.5669 \Delta (k_{\mathrm{Fx}}) \label{reduced3}.
\end{align}
Here $T_c$ is the critical temperature, calculated according to each gap model parametrization, and $T$ is the local temperature of the core, $T = \tilde{T} \exp(-\phi(r))$, where $\tilde{T}$ is the temperature of the isothermal core as seen at infinity. The heat capacity of neutrons and protons is similarly reduced when they are superfluid or superconducting, $\mathrm{C}_{p,n}^{\mathrm{superfluid}} = \mathrm{C}_{p,n} ~ R$.

Equations (\ref{reduced1})--(\ref{reduced3}) are simplifications of the results of full calculations, that find the proper quantum state rearrangements in the energy levels of the matter once neutron and proton pairing are taken into account \cite{Yakovlev}. We acknowledge the difference in the reduction rates between the full calculation and the approximations (Eq. (\ref{reduced1})--(\ref{reduced3})), especially at $T \ll T_c$ and when superconductivity and superfluidity are simultaneously present in the star's core. This difference becomes large only well-below the critical temperature $T_c$, where the neutrinos are already suppressed, thus our results are not expected to change qualitatively. Results of full calculations will be explored in a forthcoming paper.

To characterize the different gap models, we work with the polynomial parametrizations for the superfluid gap\cite{Hoetal}
\begin{equation}\label{polynomial}
\Delta\left(k_{\mathrm{Fx}}\right)=\Delta_{0} \frac{\left(k_{\mathrm{Fx}}-k_{0}\right)^{2}}{\left(k_{\mathrm{Fx}}-k_{0}\right)^{2}+k_{1}} \frac{\left(k_{\mathrm{Fx}}-k_{2}\right)^{2}}{\left(k_{\mathrm{Fx}}-k_{2}\right)^{2}+k_{3}},
\end{equation}
where $\Delta_{0}$, $k_{0}$, $k_{1}$, $k_{2}$ and $k_{3}$ are free parameters fitted to adjust the amplitudes and widths of each gap model. The resulting superfluid gaps are shown in Fig. \ref{parametrizations} and their parameters can be found in Ref. \citenum{Hoetal}.

\section{Results}\label{results}

\subsection{Direct Urca cooling and the dependence on gap model}

\begin{figure}[ht]
    \begin{center}
	\includegraphics[width=\columnwidth]{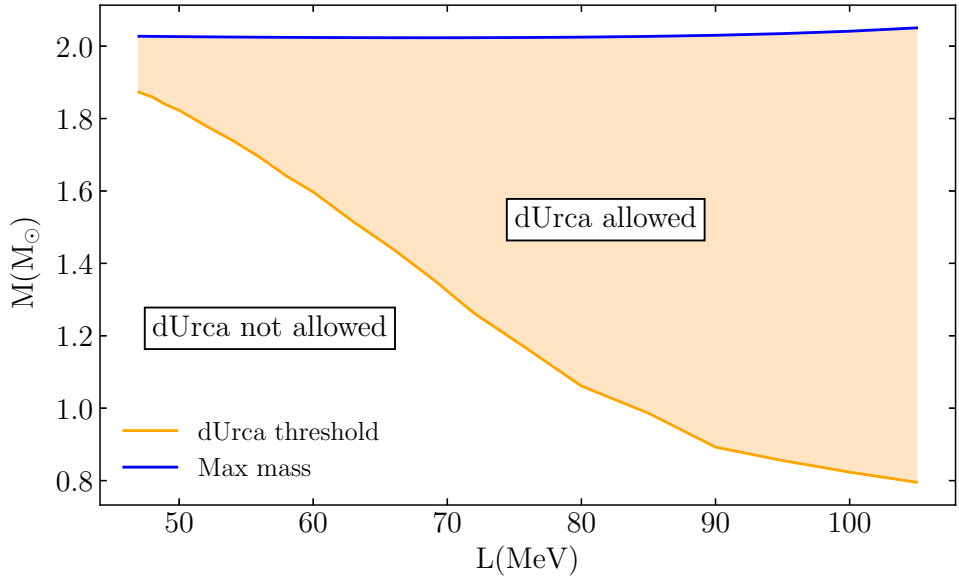}
	\end{center}
	\caption{Direct Urca threshold calculation (orange curve) for all EOS without nuclear pairing. The shaded area corresponds to the phase space where direct Urca reactions are allowed, the maximum mass reached for each EOS is also shown (blue curve).}
	\label{DURCA threshold}
\end{figure}

We first calculated neutron star models without the effects of pairing, to determine when direct Urca reactions are kinematically allowed, as a function of the neutron star mass $M$ and slope of the symmetry energy $L$. The direct Urca threshold for all EOS is shown in Fig. \ref{DURCA threshold}. All stars with masses above the $\mathrm{M}_{\mathrm{dUrca}}$ curve can have part or the entirety of their core emitting neutrinos through direct Urca reactions. For low $L$ EOS, only stars with large masses $(M \geq 1.8 M_{\odot})$ have a proton fraction large enough to allow direct Urca. On the right hand side of the plot, high $L$ EOS, which have a larger proton fraction, can accommodate direct Urca reactions even for very low mass neutron stars $(M \leq 1.0 M_{\odot})$. Note that we include neutron stars with $M < 1.0 M_{\odot}$ to show the full parameter space, although it is not expected that such low mass neutron stars are formed in reality.

\begin{figure}[ht]
    \begin{center}
	\includegraphics[width=\columnwidth]{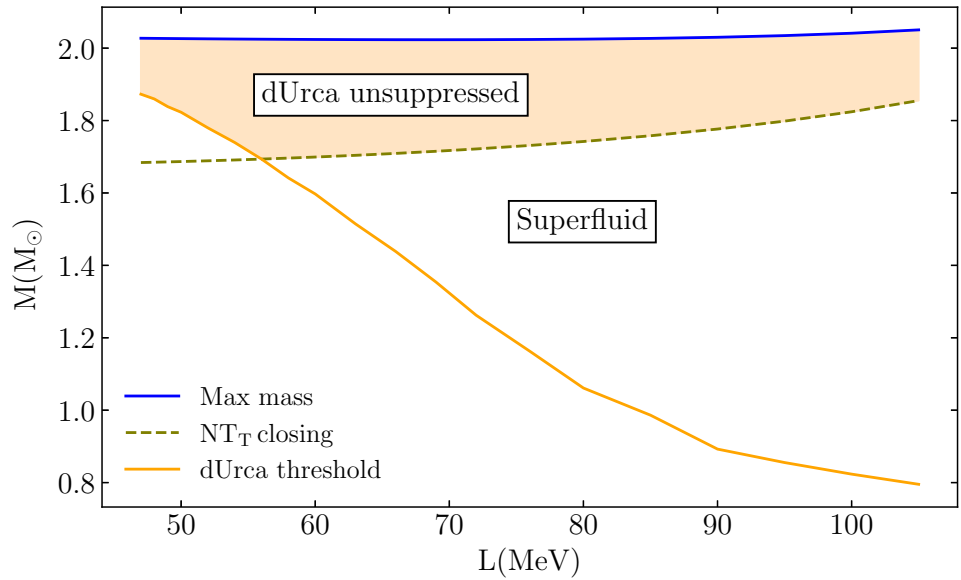}
	\end{center}
	\caption{Direct Urca threshold calculation (orange curve) and closing curve of superfluid gap model $\mathrm{NT}_\mathrm{T}$ (dashed olive). The area between the orange and dashed olive lines corresponds to exponential suppression of direct Urca reactions. Above the dashed line, there is no suppression.} 
	\label{nt_t}
\end{figure}

\begin{figure}[ht]
    \begin{center}
	\includegraphics[width=\columnwidth]{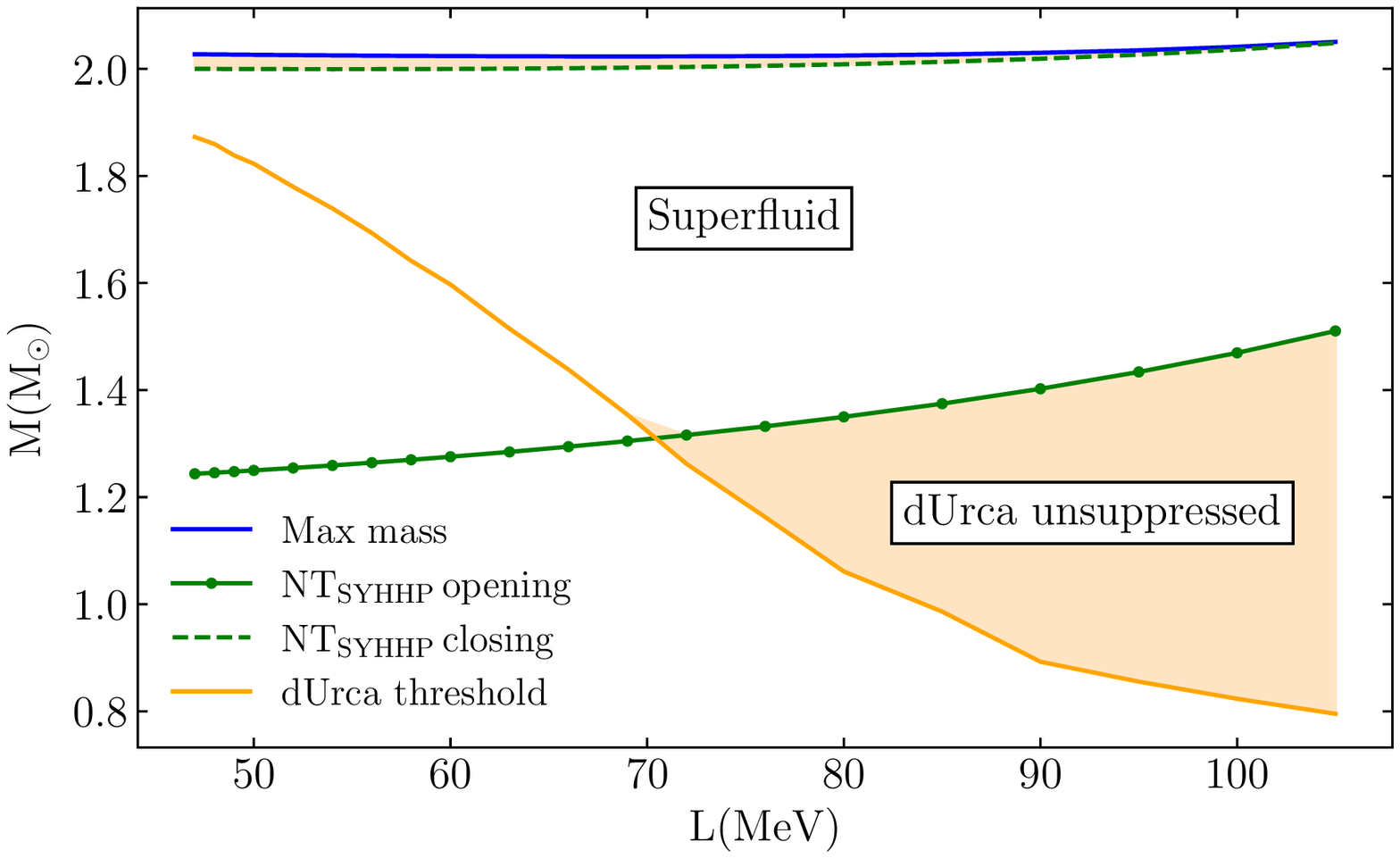}
	\end{center}
	\caption{The direct Urca threshold (orange curve), and the opening (dotted green) and closing (dashed green) curves of superfluid gap $\mathrm{NT}_\mathrm{SYHHP}$. The area bounded by the orange, the dotted green and the dashed green lines corresponds to exponential suppression of direct Urca reactions. In all highlighted areas, there is no suppression.}
	\label{nt_syhhp}
\end{figure}

The range of parameter space in which direct Urca cooling can occur may change once we include superfluidity, as a consequence of the exponential suppression of direct Urca rates in the presence of Cooper pairs. Depending on the range of density over which superfluidity occurs, there can be different effects on the direct Urca parameter space. For example, shown in Fig. \ref{nt_t}, the gap model $NT_{\rm T}$ predicts that for $L \geq 60 \, \mathrm{ MeV}$, part of the core of all neutron stars with masses $M \leq (1.7$--$1.8) \, M_{\odot}$ will be superfluid, thus reducing their total neutrino luminosity. Therefore, owing to the combination of the direct Urca threshold and the suppression by superfluidity, in this case only high mass stars $M \geq (1.7$--$1.8) \, M_{\odot}$ can significantly cool through direct Urca reactions, even for large values of $L$. In all figures where superfluidity gap models are taken into consideration, we also plot their closing curves, which are obtained by finding the densities for which the predicted gap model $T_c (\rho)$ matches the star's core temperature $T$, for $\tilde{T}=2.5 \times 10^7 \mathrm{K}$, the inferred core temperature for MXB~1659-29 \cite{Brownetal}.

Another interesting situation is when there is a late opening of the superfluid gap, as in the case of $NT_{\rm SYHHP}$, Fig. \ref{nt_syhhp}. In this situation, for large $L$ $(L \geq 70 \, \mathrm{MeV})$ and low mass stars $(M \leq 1.3 M_{\odot} - 1.4 M_{\odot})$, direct Urca reactions are allowed to happen because there is no superfluidity in the core yet. Cooper pair formation happens for larger mass stars or lower $L$, when direct Urca is suppressed, until there is no more superfluidity, that is, the gap closes, at higher masses. Therefore, direct Urca reactions are allowed again, close to maximum mass stars. In this situation, there are two regions of the phase diagram where direct Urca processes are significant: large $L$ and low mass stars or very high mass stars $(M \approx 2.0 M_{\odot})$ for all $L$.

There are also cases in which the superfluid part of the core is in a region where direct Urca reactions do not happen. Then, superfluidity does not change the calculations of neutrino luminosity shown in Fig. \ref{DURCA threshold}. This is the case for gap models $\mathrm{NT}_{\mathrm{EEHOr}}$, $\mathrm{PS}_{\mathrm{BS}}$ and $\mathrm{PS}_{\mathrm{CCYps}}$. The last possible situation is represented by gap models $\mathrm{NT}_{\mathrm{AO}}$, $\mathrm{NT}_{\mathrm{BEEHS}}$ and $\mathrm{NT}_{\mathrm{TToa}}$, which predict that the whole core of the neutron star is superfluid, severely suppressing direct Urca for all EOS and all masses.  

\subsection{Application to MXB~1659-29}

We now discuss the predictions of our model for the accreting neutron star MXB~1659-29. As described in Ref. \citenum{Brownetal}, knowing the total mass accreted onto the neutron star during an outburst, one can estimate the total energy deposited in the crust. Assuming this energy is almost completely conducted into the core, and, after the core reaches thermal equilibrium, radiated away by neutrinos, we get an estimate of the star's neutrino luminosity. For the source MXB~1659-29, this luminosity is estimated to be $L_{\nu} \approx 3 \times 10^{27}\ \mathrm{J}/\mathrm{s} = 3 \times 10^{34}\ \mathrm{erg}/\mathrm{s}$. 

This estimate was shown \cite{Brownetal} to be insensitive (less than a factor of two) to variations in accretion rate during outburst and the outburst recurrence time, neutron star mass or EOS, for a core temperature of $\tilde{T} = 2.5 \times 10^7 \ \mathrm{K}$ and a light element envelope composition. As detailed in Ref. \citenum {Brownetal2017}, the envelope composition is a crucial parameter in modelling the relationship between the internal temperature of the star and its surface temperature. For this source, a heavy element envelope would correspond to a core temperature of $\tilde{T} = 5 \times 10^7\ \mathrm{K}$, which increases the star's predicted total neutrino luminosity by two orders of magnitude, however, this scenario is inconsistent with the measured accretion rates. Ref. \citenum{Brownetal} also found that the envelope's impurity parameter and distance uncertainties can change the inferred luminosity above by an additional factor of $\lesssim 2$. For this study, we will take the value $L_{\nu} = 3 \times 10^{34}\  \mathrm{erg}/\mathrm{s}$ as the correct value of neutrino luminosity. 

\begin{figure}[ht]
\begin{center}
\includegraphics[width=0.8\columnwidth]{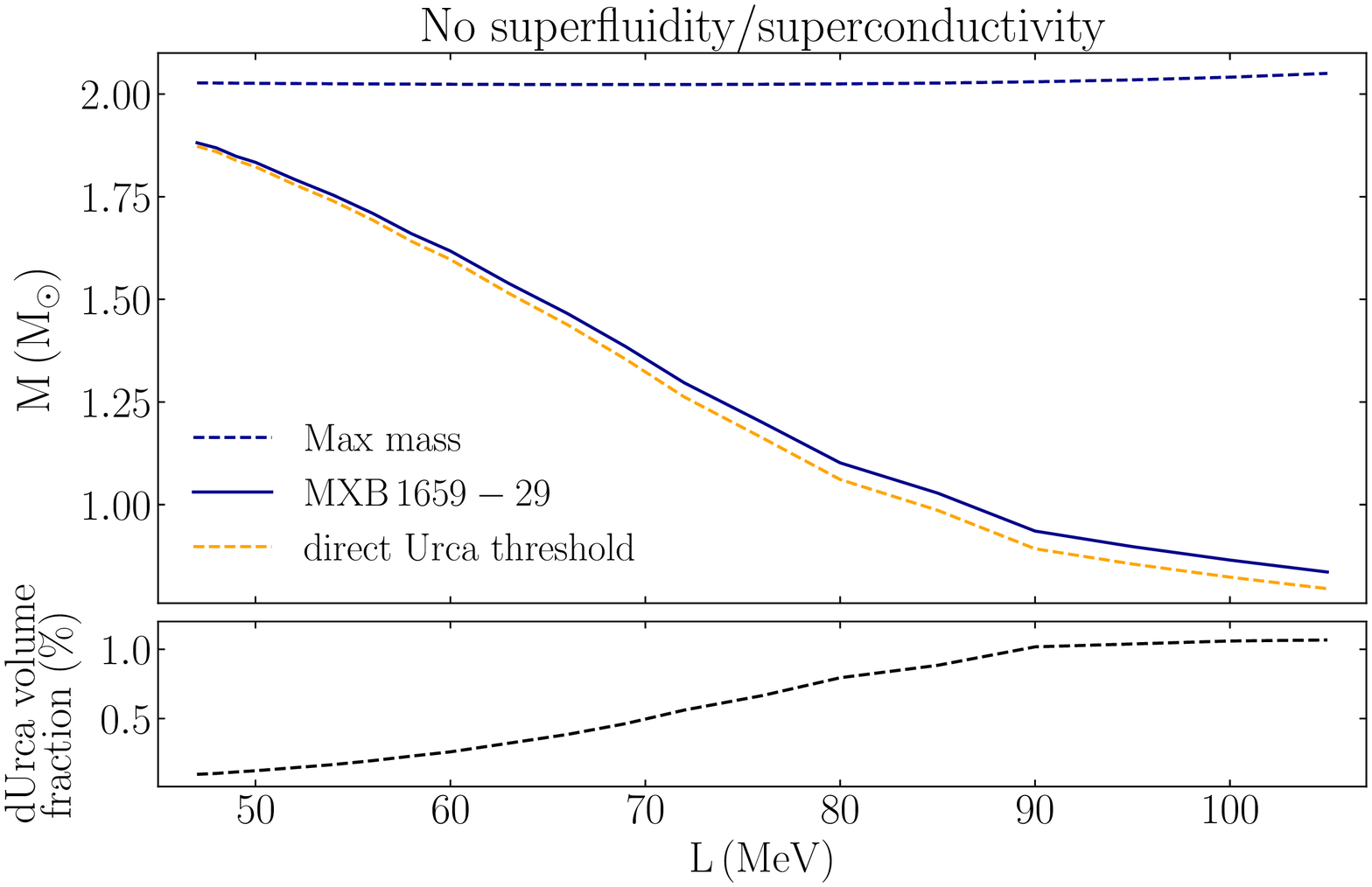}
\includegraphics[width=0.7\columnwidth]{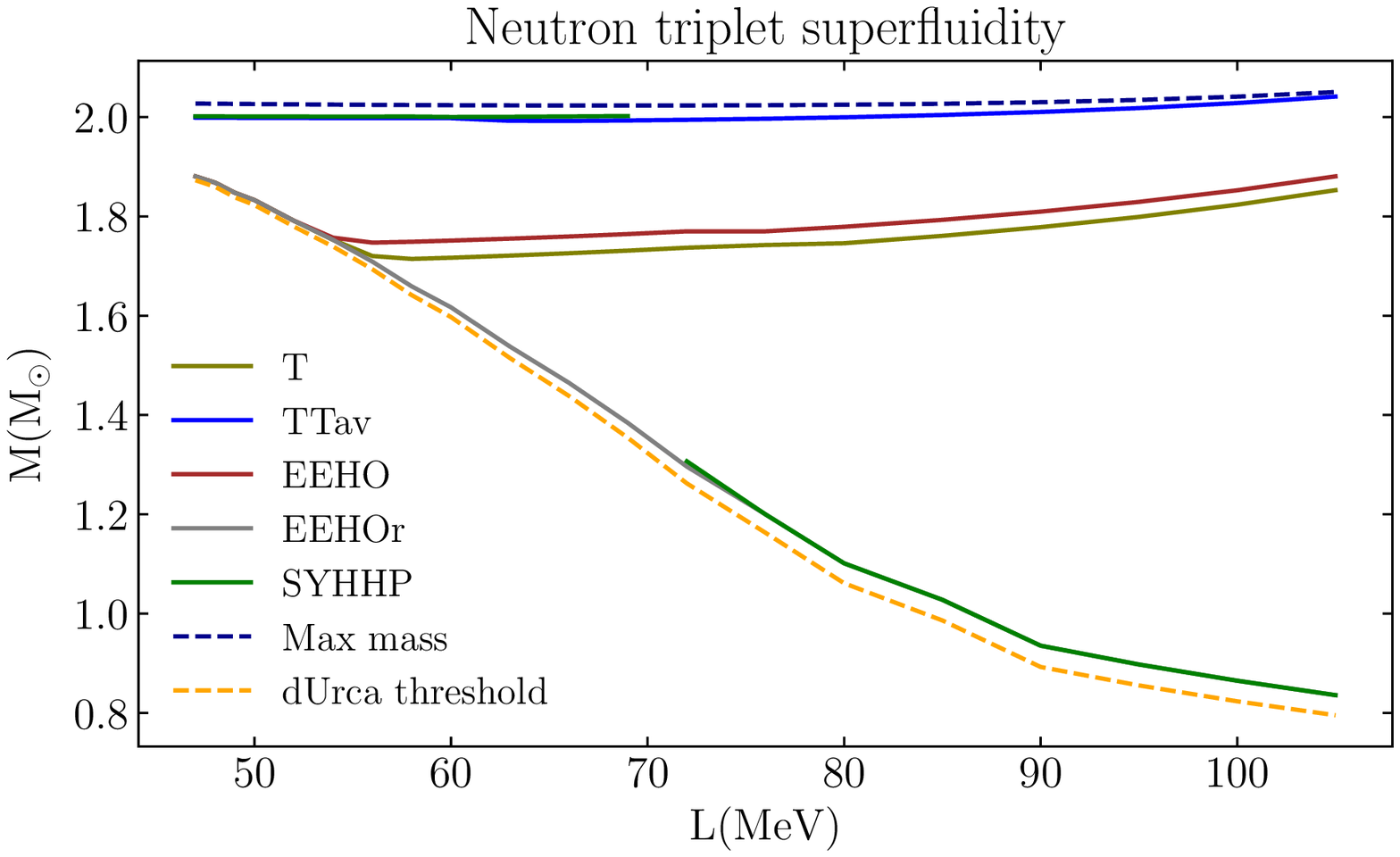}
\includegraphics[width=0.7\columnwidth]{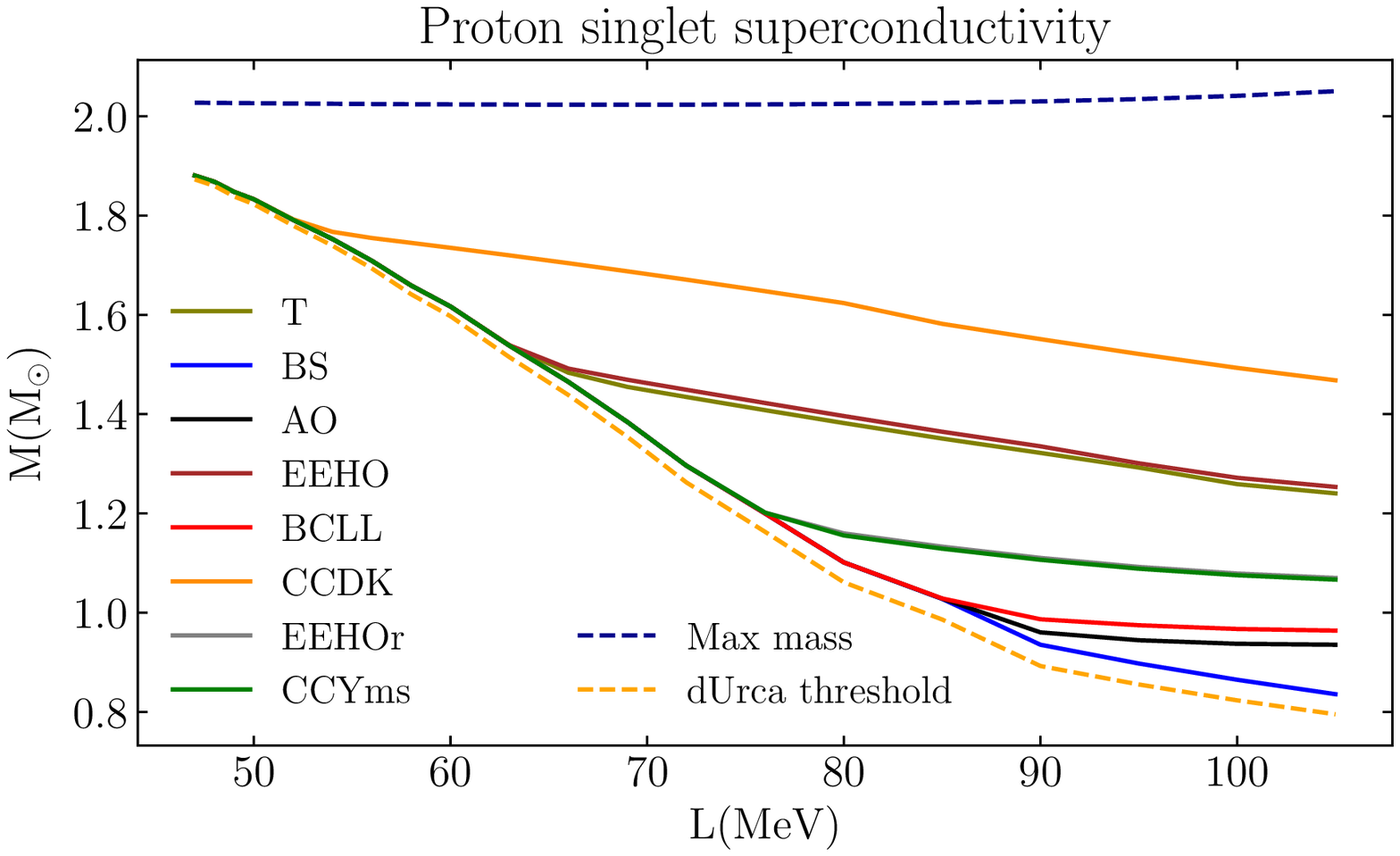}
\caption{Masses of stars with luminosity $L_{\nu} = 3 \cdot 10^{27} \mathrm{J}/\mathrm{s}$, corresponding to MXB~1659-29 source, for all EOS. 
	(a) No superfluidity/superconductivity (dark blue curve). The bottom panel displays the percentage of the core volume involved in unsupressed direct Urca reactions. (b) For all gap models of neutron triplet superfluidity. The curve displayed in panel a is under the gap model $\mathrm{NT}_{\mathrm{EEHOOr}}$ curve. 
	(c) For all gap models of proton singlet superconductivity. Gap models $\mathrm{PS}_{\mathrm{CCYps}}$ and $\mathrm{PS}_{\mathrm{BS}}$ predict the same curve.  
	}\label{L3e27}
\end{center}
\end{figure}

To attempt to match the inferred neutrino luminosity of the source MXB~1659-29, we calculate the masses of stars with that total luminosity, for all combinations of EOS and gap models. The results are shown in Fig. \ref{L3e27}. Panel (a) displays only the no superfluidity case, whereas panel (b) contains the predictions for neutron triplet superfluidity and panel (c), for proton singlet superconductivity. In all cases, we observe that only a small fraction of the volume of the core, from $0.10 \%$ to $1.07 \%$ depending on the EOS, is involved in unsuppressed direct Urca cooling. These volume fractions are shown in the bottom panel of panel (a). 

\begin{figure}[ht]
	\begin{center}
		\includegraphics[width=\columnwidth]{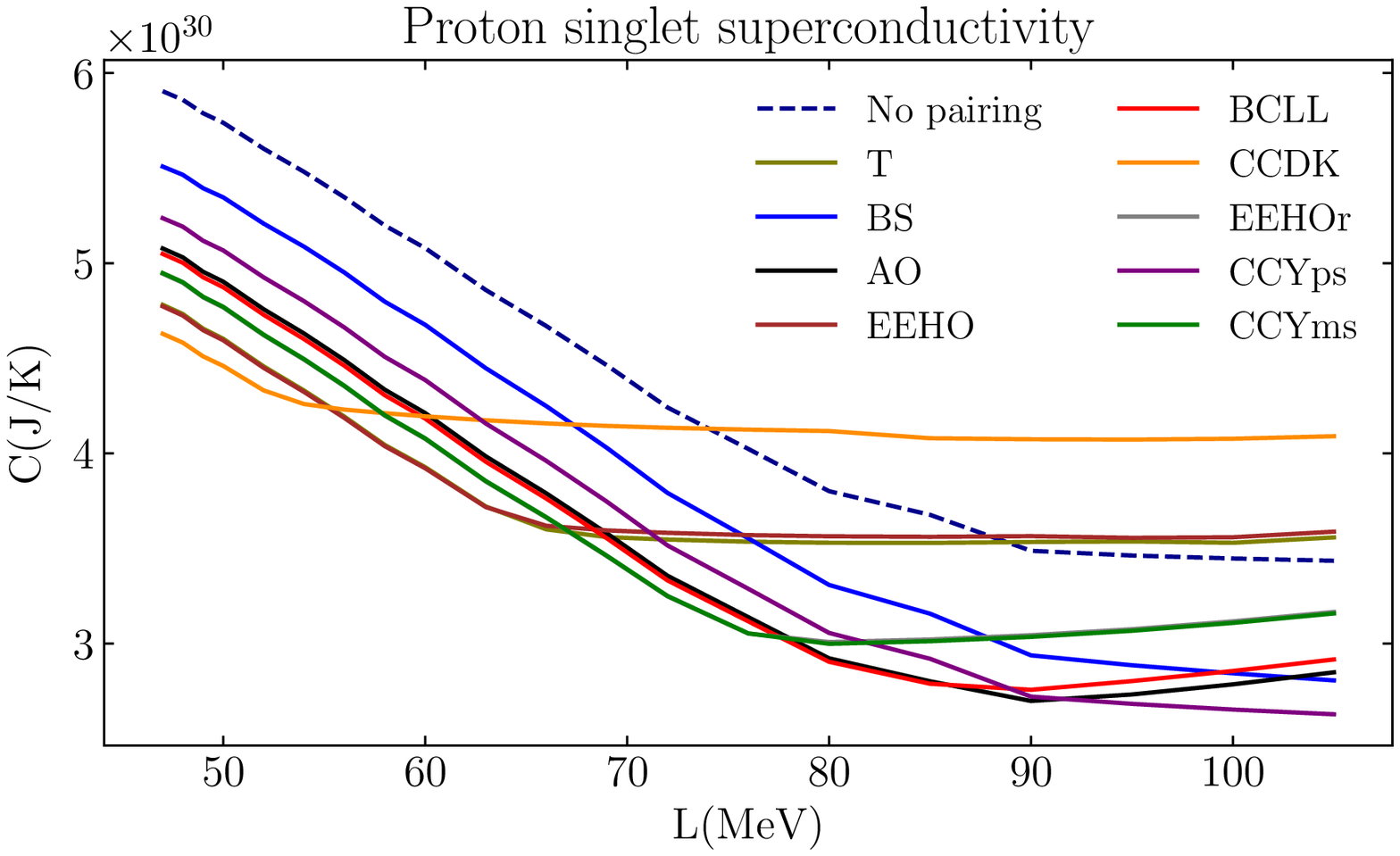}
        \includegraphics[width=\columnwidth]{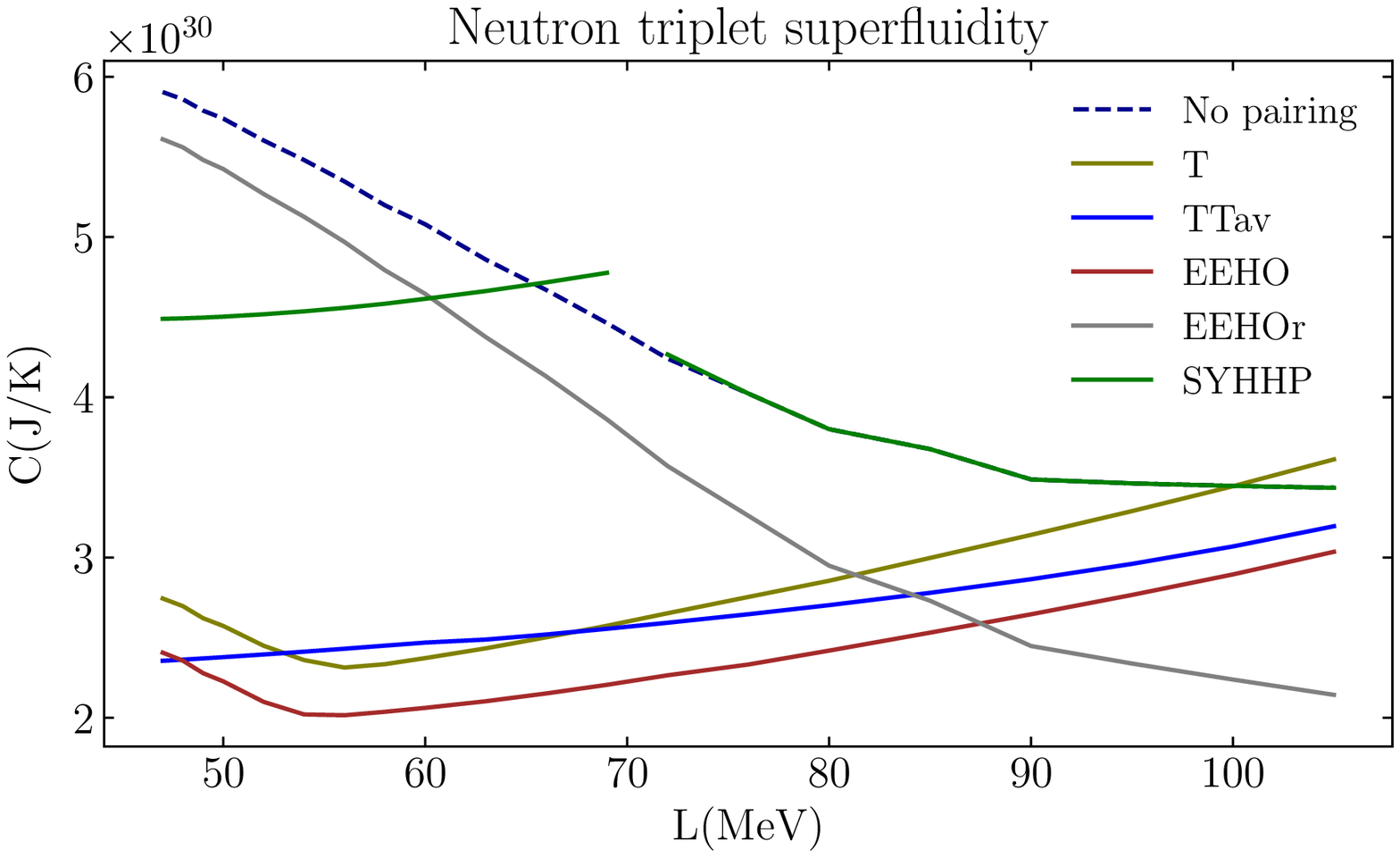}
\caption{Total heat capacity $\mathrm{C} ~ (\mathrm{J}/\mathrm{K})$ versus $L ~ (\mathrm{MeV})$ for stars with luminosity $L_{\nu} = 3 \cdot 10^{27} \mathrm{J}/\mathrm{s}$ for all EOS studied here.
	(a) Proton singlet superconductivity (b) Neutron triplet superfluidity.}
\label{heatcap1}
\end{center}
\end{figure}

\begin{figure}[ht]
	\begin{center}
		\includegraphics[width=\columnwidth]{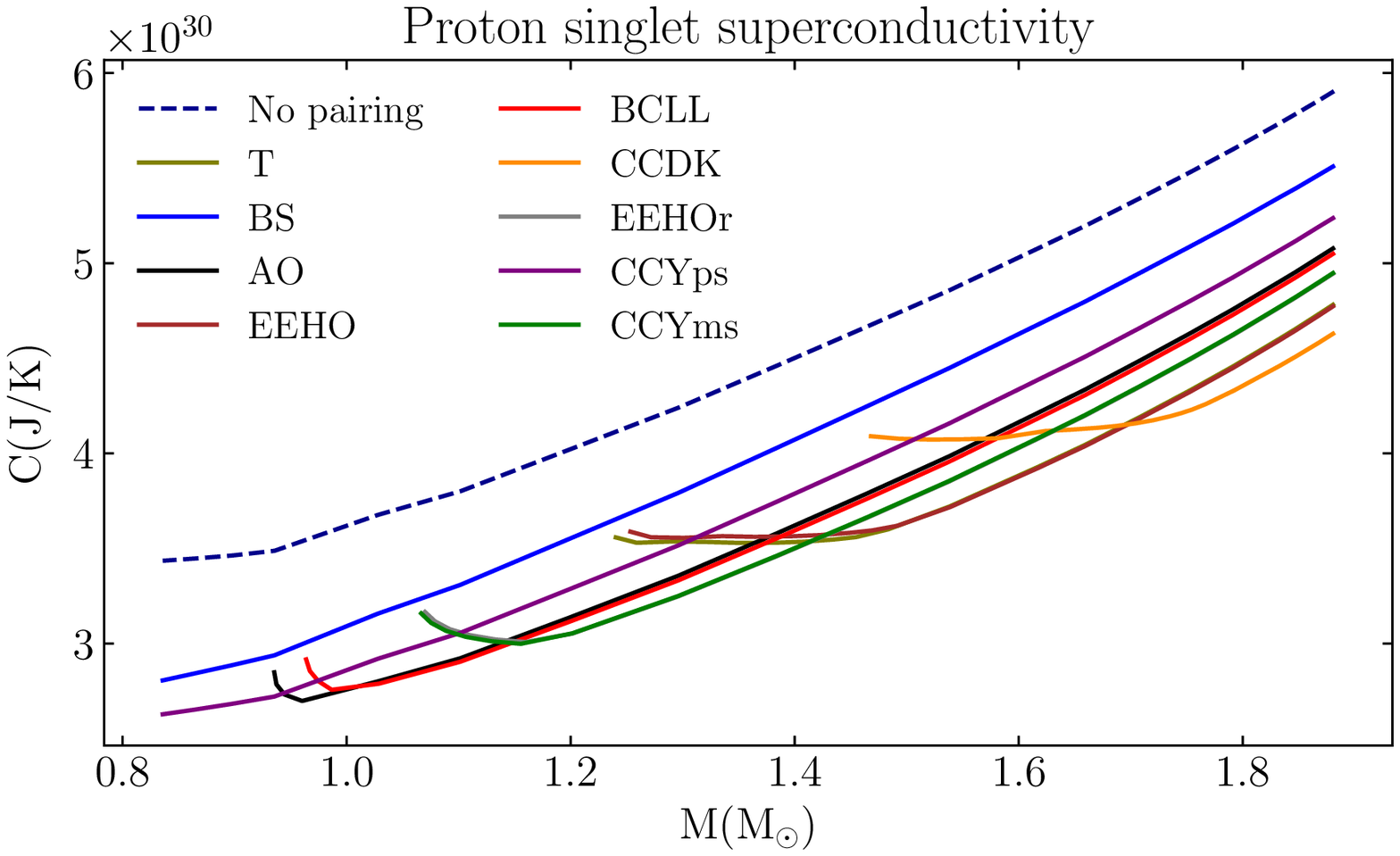}
		\includegraphics[width=\columnwidth]{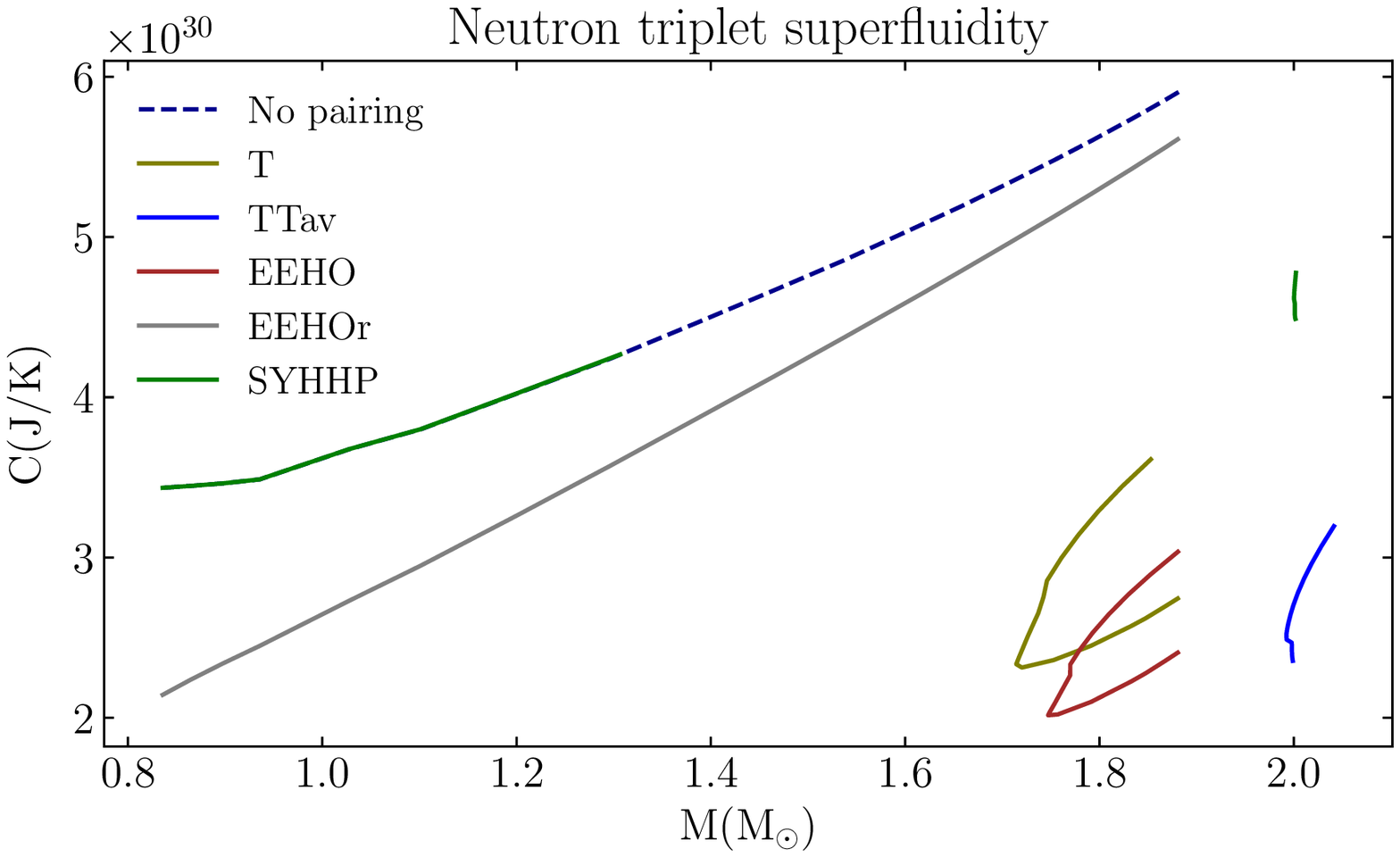}
\caption{Total heat capacity $\mathrm{C} ~ (\mathrm{J}/\mathrm{K})$ versus total mass $\mathrm{M} ~ (\mathrm{M}_{\odot})$ for stars with luminosity $L_{\nu} = 3 \cdot 10^{27} \mathrm{J}/\mathrm{s}$ for all EOS studied here.
	(a) Proton singlet superconductivity (b) Neutron triplet superfluidity.}
\label{heatcap2}
\end{center}
\end{figure}

We also calculated the total neutron star heat capacity, for all combinations of EOS and gap models. The results are shown in Fig. \ref{heatcap1}. Note that larger mass stars have larger heat capacities, which is a trend one observes in the figure above and in Fig. \ref{heatcap2}. Specifically, for proton superconductivity, high $L$ EOS have lower mass stars achieving the observed luminosity of $L_{\nu} = 3 \times 10^{34}\  \mathrm{erg}/\mathrm{s}$, thus Fig. \ref{heatcap1}a shows a general trend of lower heat capacities for higher $L$s. The opposite trend happens for neutron superfluidity, where most heat capacity curves increase with increasing $L$, reflecting that higher mass stars are produced in that region. The C-shaped curves on the right bottom of Fig. \ref{heatcap2}b are a consequence of this phenomenon. Note that different $L$ EOS can generate stars with the same mass and luminosity, for a given gap model. The fact that we are not comparing same size stars on Fig. \ref{heatcap1} explains why some heat capacity curves of superfluid stars overcome the heat capacity curve of stars without nuclear pairing. This situation does not happen in Fig. \ref{heatcap2}.

\section{Discussion}\label{discussion}

Comparing the predictions of each gap model and EOS combination with the inferred neutrino luminosity of the accreting neutron star MXB~1659-29 (Fig. \ref{L3e27}), we find that three models are disfavoured and can be excluded. $\mathrm{NT}_{\mathrm{AO}}$, $\mathrm{NT}_{\mathrm{BEEHS}}$ and $\mathrm{NT}_{\mathrm{TToa}}$ predict that the whole core of the star will be superfluid at the core temperature of MXB~1659-29, hence direct Urca reactions are completely suppressed and we are not able to reproduce the inferred neutrino luminosity of MXB~1659-29. Other fast cooling processes would be equally suppressed by superfluidity, such that these models describe slow cooling stars for any EOS, thus, are inconsistent with the data. These gap models have similar opening and closing densities, but different amplitudes of the critical temperature $T_c$, suggesting that their location and width are the determinant factors in the luminosity prediction. This result is a consequence of the low temperature of MXB~1659-29's core in comparison with the models' $T_c$ (Fig. \ref{parametrizations}). Therefore, we expect that any gap models with similar opening and closing densities will be unable to reproduce the data.  

The other gap models can reproduce the observed neutrino luminosity. However, gap models $\mathrm{NT}_{\mathrm{EEHOr}}$, $\mathrm{PS}_{\mathrm{BS}}$ and $\mathrm{PS}_{\mathrm{CCYps}}$ are disfavoured, particularly for high $L$ EOS. These models close before the onset of direct Urca for all EOS, hence their predictions for neutrino luminosity are the same as in the case without superfluidity. Their calculated star masses are very close to the direct Urca threshold masses, and for high $L ~(L \geq 85 ~ \mathrm{MeV})$, they predict that all neutron stars will be fast cooling, which is in disagreement with observations. For low $L$, their predictions agree with data if less than $0.5 \%$ of the core volume of the star is involved in unsuppressed direct Urca reactions, which suggests fine-tuning because it requires that the neutron star mass happens to lie within a narrow range of masses near the dURCA threshold.

Of the remaining gap models, we highlight $NT_{\rm SYHHP}$, which predicts that only very high mass stars have unsuppressed direct Urca for low and intermediate $L$ $(L \leq 70 \, \mathrm{MeV})$, whereas for high $L$, both low mass stars and very high mass stars cool through this fast process. This superfluid model is the only one predicting direct Urca cooling in low mass stars $(M \leq 1.3 \, M_{\odot})$, thus, a future observation consistent with this situation would favour this model and $L \geq 70 \, \mathrm{MeV}$. At the same time, it is able to accommodate slow cooling for stars with masses $2.0 \, M_{\odot} \geq M \geq 1.3 \, M_{\odot}$, in agreement with luminosity data from other transiently-accreting and isolated sources, for example, the ones studied in Ref. \citenum{Beloinetal2019, Pageetal2009}.

Proton superconductivity models $\mathrm{PS}_{\mathrm{AO}}$, $\mathrm{PS}_{\mathrm{BCLL}}$, $\mathrm{PS}_{\mathrm{EEHOr}}$ and $\mathrm{PS}_{\mathrm{CCYms}}$ also predict fast cooling for low mass stars with $L \geq 70 \, \mathrm{MeV}$, but, differently from the superfluid model $NT_{\rm SYHHP}$, by themselves they do not accomodate slow cooling for stars with masses $M \geq 1.3 \, M_{\odot}$, thus are inconsistent with data from other sources. However, a more realistic description of a neutron star would include a combination of neutron superfluidity and proton superconductivity, which can significantly change the results shown here. Therefore, the proton superconductivity models above can not be excluded under this argument. 

\begin{figure}[ht]
	\begin{center}
		\includegraphics[width=\columnwidth]{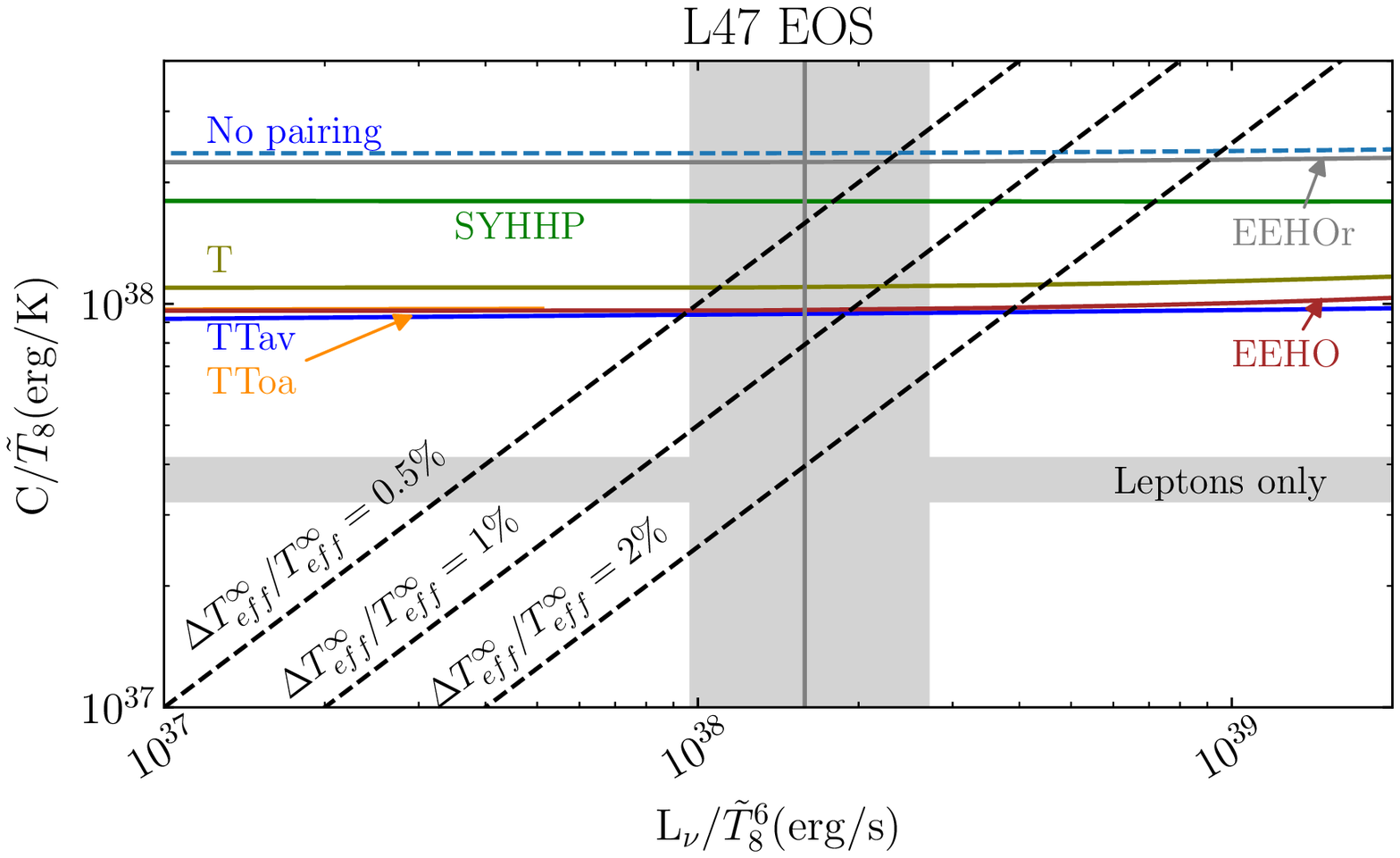}
        \includegraphics[width=\columnwidth]{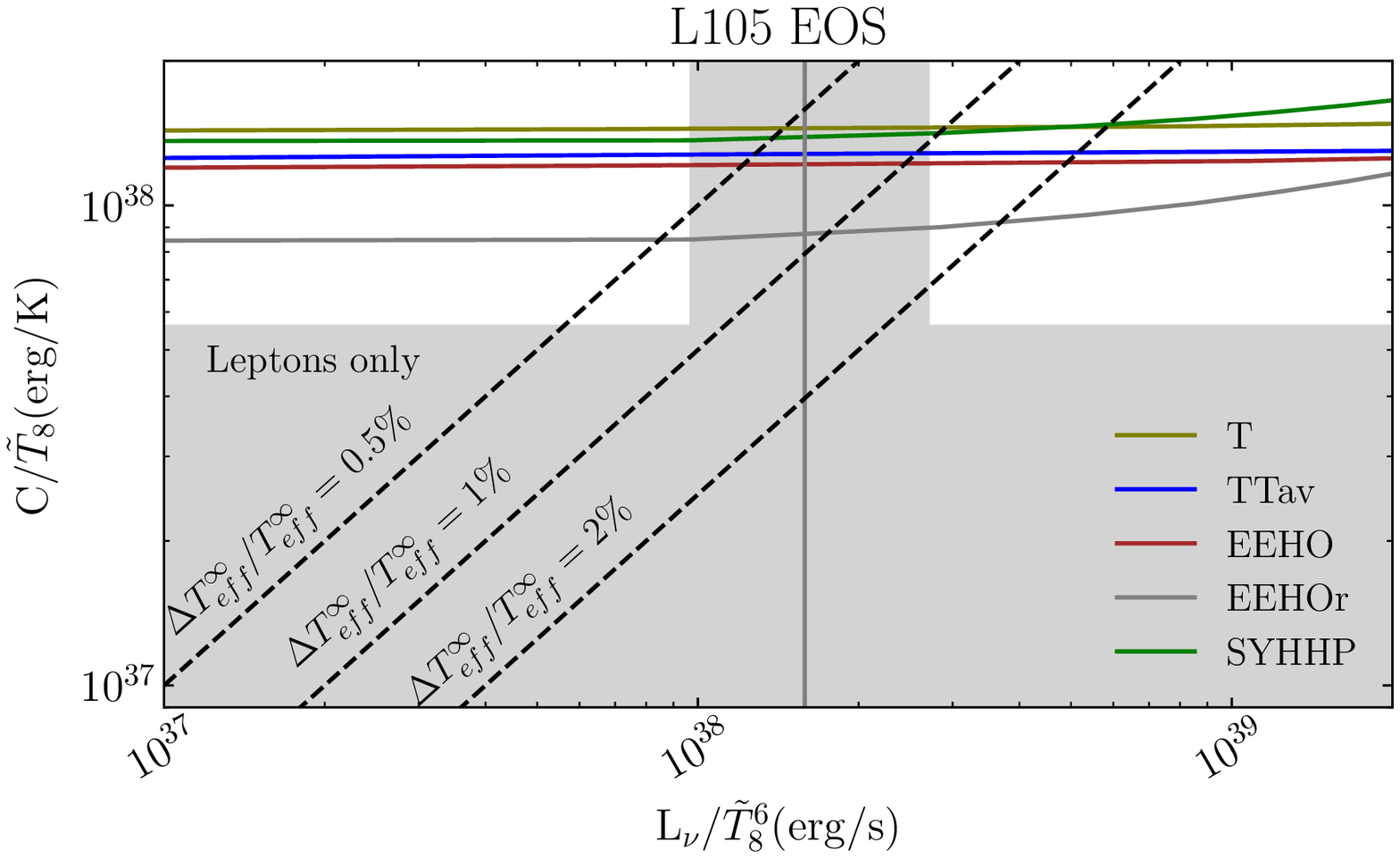}
\caption{Total heat capacity $\mathrm{C}/\tilde{T}_8 ~ (\mathrm{erg}/\mathrm{K})$ versus neutrino luminosity $\mathrm{L}_{\nu}/\tilde{T}_8^6 ~ (\mathrm{erg}/\mathrm{s})$ for neutron triplet gap models studied here. $\mathrm{C}_\mathrm{leptons}$ is the lower limit of only electrons and muons contributing, mimicking a completely superfluid star. The vertical line and grey region around it correspond to the inferred luminosity of $\mathrm{L}_{\nu}= 3 \times 10^{34}\ \mathrm{erg}/\mathrm{s}$ and uncertainty for source MXB~1659-29, more details on Ref. \cite{Brownetal}. The diagonal lines $\Delta \mathrm{T}_{\mathrm{eff}}^{\infty}/\mathrm{T}_{\mathrm{eff}}^{\infty}$ indicate temperature variation of the star's effective temperature after a period of ten years.
	(a) $L47$ EOS (b) $L105$ EOS}
\label{heatcapcomparison}
\end{center}
\end{figure}

We reproduce Ref. \citenum{Brownetal}’s Fig. 3 here, adding the calculated total heat capacities of a star for each combination of EOS and gap model. For better visualization, only neutron triplet results for two EOS are shown in Fig. \ref{heatcapcomparison}. We show that the difference between heat capacities for each case is a factor of 2 at most, thus, one needs a few percent precision in observations of temperature variation to discriminate between particle composition scenarios \cite{Brownetal}.

Knowing $L$ with precision can eliminate or favour certain gap models. However, the current predictions from gravitational waves \cite{Abbott:PRL2017, Abbott:2018exr} and nuclear experiments, such as PREX-II \cite{Reed:2021nqk} are in tension, thus, unable to set reliable constraints on $L$. For this reason, we chose to investigate the whole phase space of $L$ for the EOS family studied here. At the moment, our results are unable to set constraints for the EOS, however, with independent measurements of luminosity and mass of a neutron star source, this goal could be achieved. In this scenario, it will be particularly relevant to include more fast cooling processes in this framework, beyond direct Urca, as well as other equations of state, potentially with exotic components like hyperons, pions or free quarks.  

We highlight that, for all cases studied here, the fraction of the core volume undergoing unsuppressed direct Urca is around $1 \%$, which suggests that neutron stars with even higher neutrino luminosities and lower core temperatures should exist. Candidates for such cold stars are the sources SAX~J1808.4-3658 and 1H~1905+000 \cite{review_accreting}. Further research is necessary to confirm whether the cooling of these sources is consistent with direct Urca reactions.  Alternatively, one can speculate whether source MXB~1659-29 is actually a neutron star with a quark core or if it cools through exotic processes, such as neutrino emission from pion or kaon condensates. In some of these cases, their emissivities could be lower by up to 3 orders of magnitude, as shown in Ref. \citenum{Yakovlev}, resulting in larger fractions of the core volume cooling through direct Urca emission.

\section{Summary}

In this paper, we studied a family of hadronic EOS with different values of the slope of the symmetry energy, $L$, combined with several models of proton superconductivity and neutron superfluidity in the neutron star core. Comparing the direct Urca neutrino luminosity calculations with data from transiently-accreting source MXB~1659-29, we can rule out gap models that predict superfluidity in the whole star's core, $\mathrm{NT}_{\mathrm{AO}}$, $\mathrm{NT}_{\mathrm{BEEHS}}$ and $\mathrm{NT}_{\mathrm{TToa}}$. We predict that pairing models with similar opening and closing densities can also be excluded. Gap models $\mathrm{NT}_{\mathrm{EEHOr}}$, $\mathrm{PS}_{\mathrm{BS}}$ and $\mathrm{PS}_{\mathrm{CCYps}}$ are able to describe the observed luminosity but are disfavoured, because, for high $L$, they predict all neutron stars will be fast cooling and for low $L$, the masses need to be close to the direct Urca threshold. In a forthcoming paper, we will investigate neutron superfluidity and proton superconductivity simultaneous presence in the star's core.

We also calculated the total heat capacities for all stars which match the inferred neutrino luminosity of MXB~1659-29. Their values are within a factor of 2, hence one needs to distinguish between $1\%$ and $2\%$ of a variation of temperature over a decade \cite{Brownetal} to differentiate between the particle composition scenarios. If independent observations of mass and luminosity can be made, they could be used to determine an accurate gap model description of the star's nuclear pairing, as well as the $L$ parameter of its EOS. Furthermore, we were able to construct several detailed realistic scenarios where MXB~1659-29's estimated luminosity was obtained, consistent with direct Urca cooling at around $1 \%$ core volume, as previously estimated. This low percentage implies that stars with even larger neutrino luminosities and colder cores should exist. Alternatively, other fast cooling processes from exotic components with a lower emissivity could be operating over a larger fraction of the core.

\section*{Acknowledgments}
We thank Sangyong Jeon for useful discussions and comments. This work was funded in part by the Natural Sciences and Engineering Research Council of Canada. M.~M.~acknowledges support by the Schlumberger Foundation Faculty for the Future Fellowship program. A.~C.~is a member of the Centre de Recherche en Astrophysique du Québec (CRAQ). M.~M.~and A.~C.~are members of the McGill Space Institute (MSI).\\

\bibliographystyle{ws-procs961x669}
\bibliography{proceedings-bib}

\end{document}